\documentclass[3p,11pt,times]{elsarticle}
\usepackage{ifpdf}
\ifpdf
  \usepackage[1.7]{bxpdfver}
\fi
\usepackage{hyperref}
\usepackage{amssymb}
\usepackage{amsmath}
\usepackage{amsfonts}
\usepackage{multirow}
\usepackage{orcidlink}

\journal{arXiv}

\begin{document}

\begin{frontmatter}

\title{A Fast and Precise Method for Searching Rectangular Tumor Regions in Brain MR Images}
\author[1]{Hidenori Takeshima \orcidlink{0000-0002-2557-8533}}
\author[2]{Shuki Maruyama \orcidlink{0000-0002-2092-4652}}

\address[1]{Advanced Technology Research Department, Research and Development Center, \\
            Canon Medical Systems Corporation,
            Tokyo, Japan}
\address[2]{Software Technologies Group, MRI Systems Development Department, MRI Systems Division, \\
            Canon Medical Systems Corporation,
            Tochigi, Japan}

\begin{abstract}

Purpose:
To develop a fast and precise method for searching rectangular regions in brain tumor images.

Methods:
The authors propose a new method for searching rectangular tumor regions in brain MR images.
The proposed method consisted of a segmentation network and a fast search method with a user-controllable search metric.
As the segmentation network, the U-Net whose encoder was replaced by the EfficientNet was used.
In the fast search method, summed-area tables were used for accelerating sums of voxels in rectangular regions.
Use of the summed-area tables enabled exhaustive search of the 3D offset (3D full search).
The search metric was designed for giving priority to cubes over oblongs, and
assigning better values for higher tumor fractions even if they exceeded target tumor fractions.
The proposed computation and metric were compared with those used in a conventional method
using the Brain Tumor Image Segmentation dataset.

Results:
When the 3D full search was used, the proposed computation (8 seconds) was 100-500 times faster than the conventional computation (11-40 minutes).
When the user-controllable parts of the search metrics were changed variously,
the tumor fractions of the proposed metric were higher than those of the conventional metric.
In addition, the conventional metric preferred oblongs whereas the proposed metric preferred cubes.

Conclusion:
The proposed method is promising for implementing fast and precise search of rectangular tumor regions,
which is useful for brain tumor diagnosis using MRI systems.
The proposed computation reduced processing times of the 3D full search, and
the proposed metric improved the quality of the assigned rectangular tumor regions.
\end{abstract}


\end{frontmatter}

\section{Introduction}

There are many methods for acquiring and analyzing brain tumor diagnosis using MRI systems\cite{MRITumorAnalysis,MRSTumorAnalysis}.
While many imaging methods can acquire large regions in sufficient spatial resolutions,
some methods such as single voxel-magnetic resonance spectroscopy (SV-MRS)\cite{MRSConsensus,MRSGABA,MRS2HG}
and magnetic resonance spectroscopic imaging (MRSI)\cite{MRSConsensus}
can acquire only small rectangular regions such as
$1.5 \times 1.5 \times 1.5$ $\text{mm}^3$,
$3 \times 3 \times 3$ $\text{mm}^3$,
and $16 \times 16 \times 1.5$ $\text{mm}^3$.
In these methods, it is essential to find rectangular regions for acquisitions.

There were several works for finding rectangular regions using
anatomical\cite{AnatomicalVOIPlacement1,AnatomicalVOIPlacement2,AnatomicalVOIPlacement3}
and
tumor\cite{TumorVOIPlacement1,TumorVOIPlacement2}
information.
In the cases of brain tumors, rectangular regions are expected to be placed appropriately within tumors.
Such rectangular regions can be placed by either
searching an optimal rectangular region using outputs of segmentation networks\cite{TumorVOIPlacement1} or
learning an end-to-end function from an image to a rectangular region\cite{TumorVOIPlacement2}.
In practical use,
the search-based methods are more attractive than the end-to-end methods
since
the segmentation networks can be trained without using ground-truth rectangular regions, and
search metrics can be controlled easily at runtime.
In the conventional search-based method\cite{TumorVOIPlacement1},
the search metric included user-controllable parameters consisting of target volume,
target tumor fraction, and their weights.

The purpose of this work is to develop a fast and precise method for searching rectangular regions in brain tumor images.
There were some drawbacks in the conventional search-based method.
First, it used 1-dimensional (1D) search for reducing processing times.
Second, it often chose oblongs rather than cubes since shapes of the rectangular regions were not user-controllable.
Third, it penalized the rectangular regions whose tumor fractions were greater than the target tumor fractions specified by the search metrics.

There were various fast algorithms included
approximate search\cite{blockmatchingsurvey}, and exhaustive search using certain
metrics\cite{BlochMatchingFFT,ActiveSearch,SummedAreaTables,IntegralImage,IntegralImageHighDim}
for searching and evaluating rectangular regions in an image.
When the exhaustive search was chosen for evaluating all candidates,
summed-area tables\cite{SummedAreaTables,IntegralImage,IntegralImageHighDim}
enabled fast and precise computation of sums in rectangular regions.

The authors propose a new method for searching rectangular tumor regions in brain tumor images.
For overcoming the first drawback of the conventional search-based method\cite{TumorVOIPlacement1},
the proposed method could 
utilize the summed-area tables
for searching a 3-dimensional (3D) offset exhaustively in practical time.
The search metric gave priority to cubes over oblongs
for overcoming the second drawback.
The third drawback was solved by
assigning better values for higher tumor fractions even if they exceeded the target tumor fractions specified by the search metric.
The preliminary works of the proposed method were published as abstracts\cite{Takeshima1,Takeshima2}.

\section{Materials and Methods}

\subsection{Overview}
\begin{figure}[t]%
\centering%
\includegraphics[width=\linewidth]{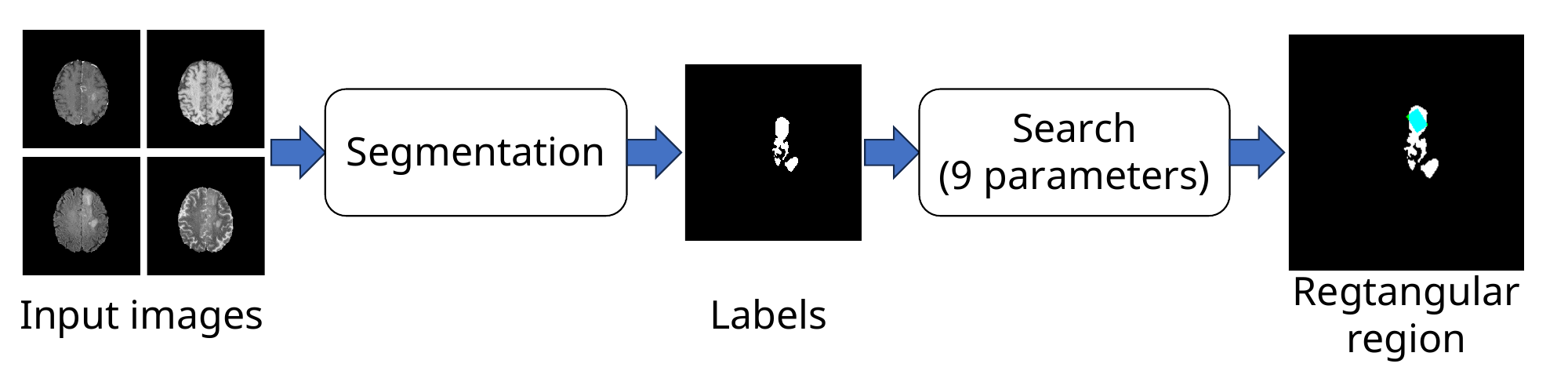}
\caption{The overview of the proposed method.
The proposed method consisted of segmentation and search steps.
In the segmentation step, all voxels were classified as tumor and non-tumor voxels.
In the search step, an optimal rectangular region was searched in 9-dimensional space
consisting of a 3D offset, a 3D size, and a 3D angle.
}\label{fig:summary}
\end{figure}

The proposed method consisted of segmentation and search steps like the conventional method\cite{TumorVOIPlacement1}.
A summary of the proposed method is shown in Fig. \ref{fig:summary}.
In the segmentation step, all voxels were classified as tumor and non-tumor voxels
using a 2-dimensional (2D) segmentation network in slice-by-slice.
In the search step, a new search method was used for finding an optimal rectangular region in 9-dimensional (9D) space
consisting of a 3D offset $V=(V_x,V_y,V_z)$, a 3D size $R=(R_x,R_y,R_z)$ and a 3D angle $\Theta=(\theta_1,\theta_2,\theta_3)$.
For computing the search metric efficiently,
the proposed method used summed-area tables\cite{SummedAreaTables,IntegralImage,IntegralImageHighDim}.
The proposed method defined a search metric as a sum of values and an adjustment function independent of 3D offsets to be searched.

\subsection{Dataset}

The open dataset originally used in the Brain Tumor Image Segmentation (BraTS)
challenge 2017\cite{BRATS1,BRATS2,BRATS3,BRATS4,BRATS5} was used for this institutional review board (IRB)-exempt study.
This dataset contained skull-stripped and co-registered multi-contrast images of brain tumors.
Their image size and spatial resolution were
$240 \times 240 \times 155$ and $1 \times 1 \times 1$ $\text{mm}^3$, respectively.
Each image consisted of
T1 weighted (T1W),
T1 weighted with contrast enhancement (T1Wc),
T2 weighted (T2W), and
fluid attenuated inversion recovery (FLAIR)
images.
In the segmentation labels, there were 4 classes consisting of 3 different tumor classes and 1 non-tumor class.

The training and validation datasets were extracted from the BraTS dataset.
The images whose qualities were visually low were manually removed.
The remaining images were split into 311 training and 87 validation images.

\subsection{Segmentation}
\begin{figure}[t]%
\centering%
\includegraphics[width=\linewidth]{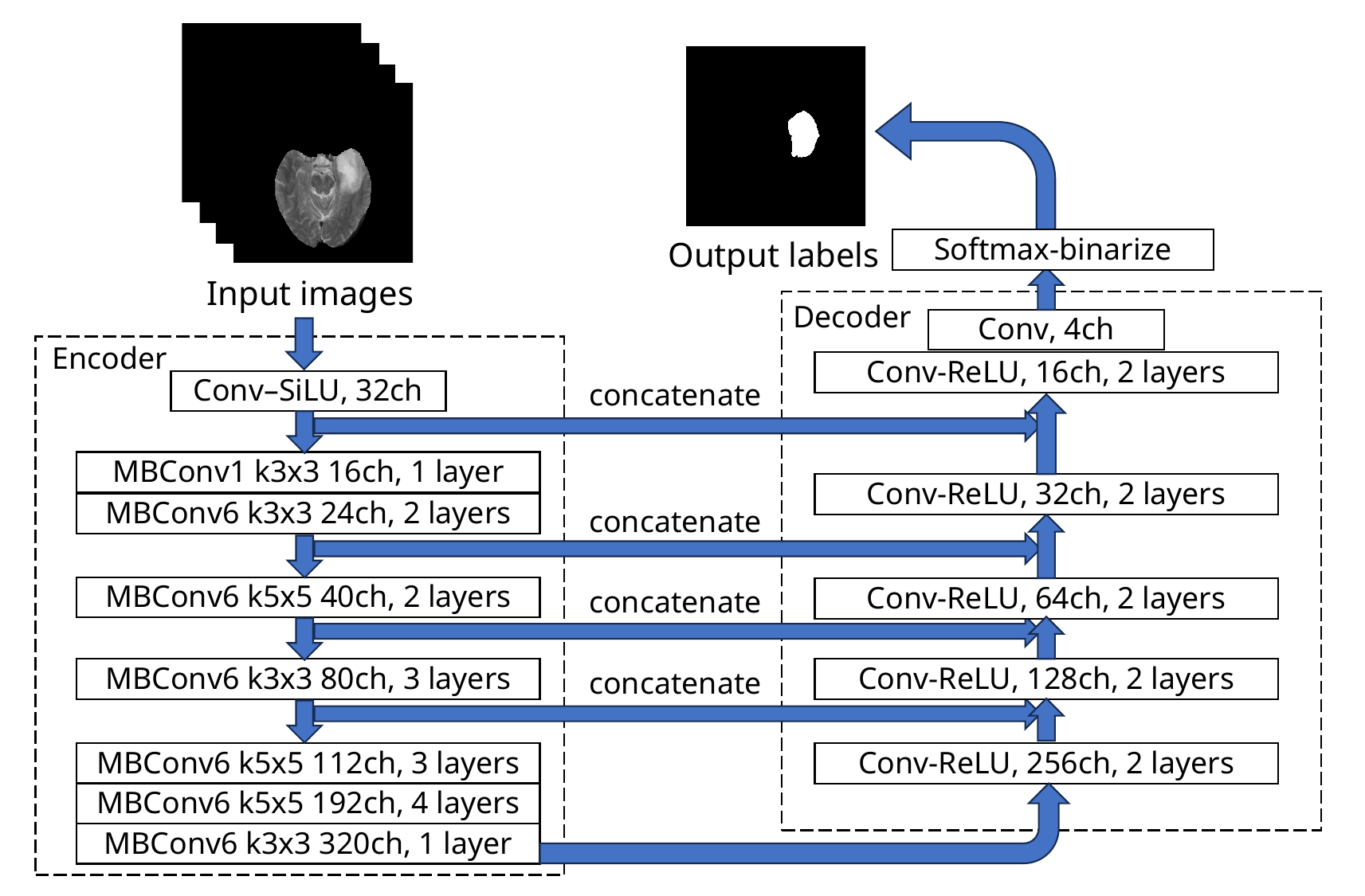}
\caption{The neural network used in tumor segmentation.
Its input images were multi-contrast images.
Its output labels were binary values consisting of 1 (tumor) and 0 (non-tumor).
The network structure was the U-Net\cite{UNet} structure
whose encoder was replaced by the EfficientNet\cite{EfficientNet}.
}\label{fig:segmentation}
\end{figure}

As the 2D segmentation networks for the proposed method,
two neural networks based on the U-Net structure\cite{UNet} was trained with two types of inputs.
In the case of the first and second networks,
the number of input channels was 4 (T1W, T1Wc, T2W, FLAIR) and 2 (T2W, FLAIR), respectively.
The number of output classes was 4 which consisted of 3 tumor and 1 non-tumor classes.
When the trained network was used in inferences, the 3 tumor classes were treated simply as a tumor class.
The segmentation models pytorch\cite{SegmentationModelsPytorch} was used for implementing
the U-Net structure whose encoder was replaced by the EfficientNet\cite{EfficientNet}.
The U-Net is an encoder-decoder network with shortcut connections from its encoder to its decoder.
The EfficientNet is a fast and accurate network originally developed for classification problems.
The encoder of the U-Net was replaced without changing the structure of the EfficientNet.
Therefore,
while the encoder used mobile inverted bottleneck convolution (MBConv)\cite{MnasNet,EfficientNet}
and sigmoid linear unit (SiLU)\cite{SiLU},
the decoder used convolution and rectified linear unit (ReLU).
A summary of the segmentation network is shown in Fig. \ref{fig:segmentation}.

To train the segmentation network, the following conditions were used.
The input images were padded from $240 \times 240$ to $256 \times 256$ for processing with the segmentation models pytorch.
The loss function was the cross-entropy loss.
The segmentation network was trained with the Adam\cite{Adam}.
The hyperparameters of the Adam were
$\beta_1=0.9$, $\beta_2=0.999$ and $\gamma=0.001$.
Other hyperparameters included
batch size of 16, and number of epochs of 50.

\subsection{Fast Computation}
\begin{figure}[t]%
\centering%
\includegraphics[width=10cm]{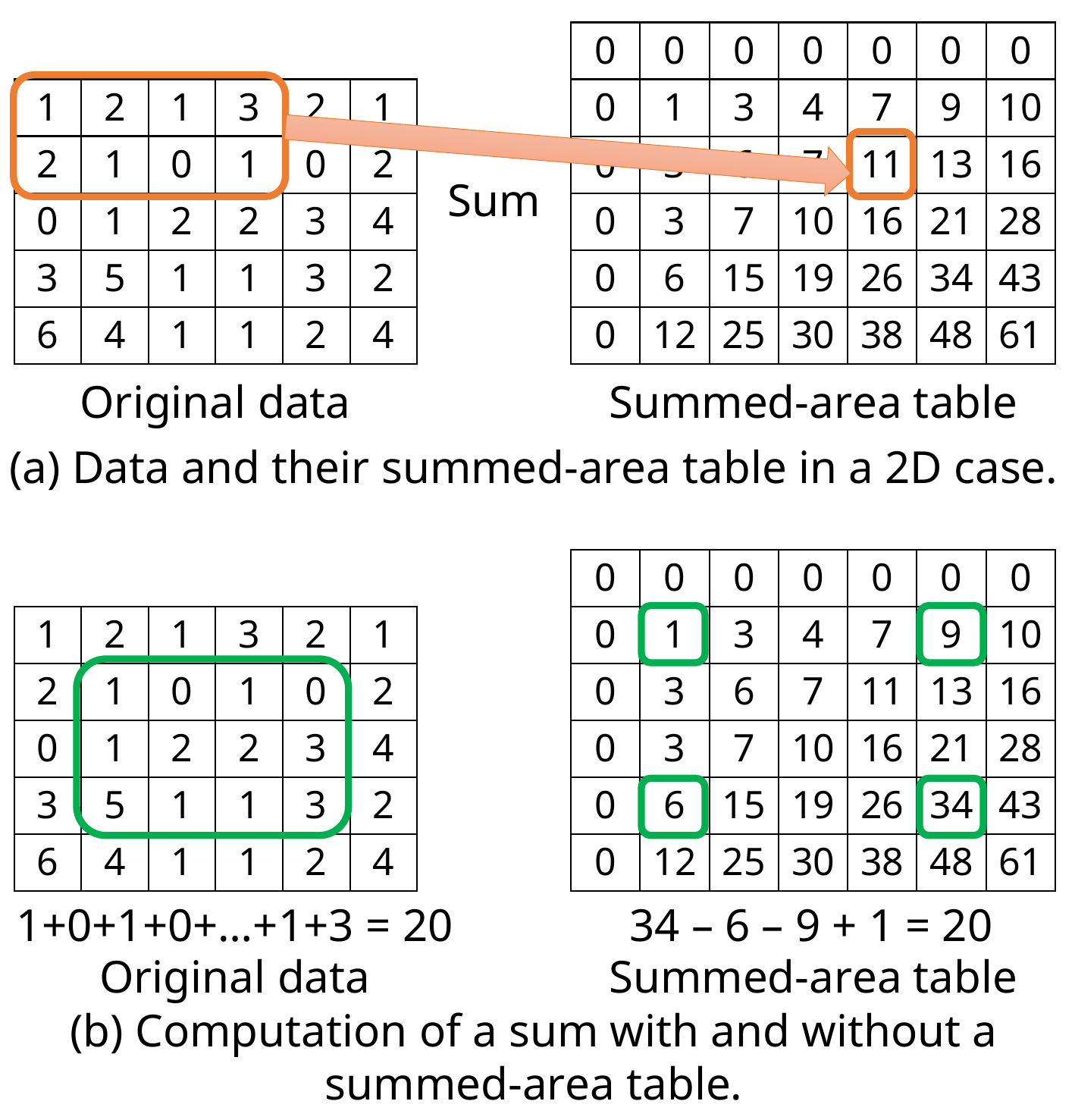}
\caption{A summed-area table in a 2D case.
As shown in (a), a summed-area table can be computed by summing rectangular regions from the left-top pixel to all pixels.
As shown in (b), a sum of a rectangular region can be computed by a fixed number of add/subtract operations
if the summed-area table is precomputed.
For the sake of conciseness, a 2D summed-area table is explained in these figures.
Actual implementation used 3D summed-area tables.
}\label{fig:summedareatable}
\end{figure}

The conventional metric\cite{TumorVOIPlacement1} $f_{conventional}$ can be represented as
\begin{equation}\label{eq:conventional}
\begin{aligned}
f_{conventional}(V,R,\Theta) =
&\exp \left( -\frac{1}{2} \left( \frac{R_xR_yR_z - l_xl_yl_z}{\sigma_R}  \right)^2  \right) \cdot \\
&\exp \left( -\frac{1}{2} \left( \frac{f_{sum}(V,R,\Theta)/(R_xR_yR_z) - f_{target}}{\sigma_f}  \right)^2  \right)
\end{aligned}
\end{equation}
where $(l_x,l_y,l_z)$ represents the target size of the rectangular region,
$\sigma_R$ and $\sigma_f$ represent user-controllable parameters,
$f_{sum}(V,R,\Theta)$ represents the sum of the tumor labels
in the region defined by the 9D parameters $(V,R,\Theta)$, and
$f_{target}$ represents the target tumor fraction.
The conventional method maximized the search metric $f_{conventional}$.
The summation part $f_{sum}(V,R,\Theta)$ can be represented as
\begin{equation}
f_{sum}(V,R,\Theta) = \sum_{z=1}^{R_z} \sum_{y=1}^{R_y} \sum_{x=1}^{R_x} S_{\Theta}(V_x+x,V_y+y,V_z+z) \label{eq:func_fsum}
\end{equation}
where $S_{\Theta}(x,y,z)$ represents the segmented label at $(x,y,z)$
in the image rotated with the angle $\Theta$.
The computational complexity was $O(R_xR_yR_z)$ for straightforward computation of $f_{sum}(V,R,\Theta)$.
The conventional method\cite{TumorVOIPlacement1} used 1D search
since the computational cost of Eq. \ref{eq:conventional} was high.

A summed-area table $T_{\Theta}(V_x,V_y,V_z)$ is a table which stores sums of values from $(1,1,1)$ to $(V_x,V_y,V_z)$ for all pixels.
Let $N_x$, $N_y$, and $N_z$ be the numbers of pixels in x, y, and z axes, respectively.
The summed-area table $T_{\Theta}(V_x,V_y,V_z)$ is defined as
\begin{equation}\label{eq:summedareatable}
T_{\Theta}(V_x,V_y,V_z) = \sum_{z=1}^{V_z} \sum_{y=1}^{V_y} \sum_{x=1}^{V_x} S_{\Theta}(x,y,z)
\end{equation}
for all pixels within
$1 \le V_x \le N_x$,
$1 \le V_y \le N_y$, and
$1 \le V_z \le N_z$.
In addition, let $T_{\Theta}(V_x,V_y,V_z)$ be $0$ if one of $V_x$, $V_y$, and $V_z$ is zero.
The computation method of a 2D summed-area table is shown in Fig. \ref{fig:summedareatable} (a).
Actual implementation used 3D summed-area tables.

Since the summed-area table $T_{\Theta}(V_x,V_y,V_z)$ depends on the angle $\Theta$,
summed-area tables are re-created whenever the angle $\Theta$ are changed.
The computational complexity is $O(V_xV_yV_z)$ for computing $T_{\Theta}(V_x,V_y,V_z)$.
By using the summed-area table $T_{\Theta}(V_x,V_y,V_z)$, the summed-area of the tumor labels can be computed as
\begin{equation}\label{eq:summedarea}
\begin{aligned}
f_{sum}(V,R,\Theta) &=   T_{\Theta}(V_x + R_x, V_y + R_y, V_z + R_z)
                - T_{\Theta}(V_x - 1, V_y + R_y, V_z + R_z)
                - T_{\Theta}(V_x + R_x, V_y -1, V_z + R_z) \\
                &- T_{\Theta}(V_x + R_x, V_y + R_y, V_z - 1)
                + T_{\Theta}(V_x - 1, V_y - 1, V_z + R_z)
                + T_{\Theta}(V_x + R_x, V_y - 1, V_z - 1) \\
                &+ T_{\Theta}(V_x - 1, V_y + R_y, V_z - 1)
                - T_{\Theta}(V_x - 1, V_y - 1, V_z - 1).
\end{aligned}
\end{equation}
The computation method of a summed-area using a 2D summed-area table is shown in Fig. \ref{fig:summedareatable} (b).
The computational complexity is $O(1)$
for computation of $f_{sum}(V,R,\Theta)$ with $T_{\Theta}(V_x,V_y,V_z)$.
By representing the number of candidates as $N$,
overall computational complexity is $O(NR_xR_yR_z)$ for the straightforward computation, and
the maximum complexity of $O(V_xV_yV_z)$ and $O(N)$
for the computation using the summed-area tables.
Therefore, use of the summed-area tables is efficient when the number of candidates $N$ is large.

The remaining part of Eq. \ref{eq:conventional} can be rewritten for evaluating individual candidates quickly.
By utilizing negative logarithmic functions and ignoring scaling factors,
Eq. \ref{eq:conventional} can be represented as
\begin{equation}\label{eq:neglog}
- \log f_{conventional}(V,R,\Theta) \propto
  \left( \frac{f_{sum}(V,R,\Theta)}{R_xR_yR_z} - f_{target} \right)^2 +
  \lambda_1 ( R_xR_yR_z - l_xl_yl_z )^2
\end{equation}
where $\lambda_1 = \sigma_R^2/\sigma_f^2$ represents the user-controllable parameter given separately in the Eq. \ref{eq:conventional}.

\subsection{Search Metric}

There were two drawbacks in the conventional metric given in Eq. \ref{eq:neglog}:
The first term of Eq. \ref{eq:neglog} penalized rectangular regions whose tumor fractions were greater than $f_{target}$, and
the second term of Eq. \ref{eq:neglog} did not penalize oblongs.

To overcome these drawbacks, the proposed method improved the search metric by
changing the first term for treating tumor fractions greater than $f_{target}$ as better, and
the second term for penalizing oblongs.
The improved search metric used in the proposed method $f_{proposed}(V,R,\Theta)$ is given as
\begin{equation}\label{eq:proposed}
\begin{aligned}
f_{proposed}(V,R,\Theta) =
  & f_{leaky}\left( f_{target} - \frac{f_{sum}(V,R,\Theta)}{R_xR_yR_z} \right) + \\
  & \lambda_2 (|R_x-l_x| + |R_y-l_y| + |R_z-l_z|)
\end{aligned}
\end{equation}
where 
$f_{leaky}$ is the leaky rectified linear unit function, and
$\lambda_2$ represents a search parameter for penalizing the shapes of rectangular regions.
The function $f_{leaky}$ is defined as
\begin{equation}
f_{leaky}(s) =
\left\{
\begin{array}{cl}
s
 & \text{if} \, w \ge 0, \text{and} \\
\beta s
 & \text{otherwise}.
\end{array}
\right.
\end{equation}
In the function $f_{leaky}$, the leaky factor $\beta > 0$ increases
the priority of the regions whose tumor fraction is greater than $f_{target}$.

\subsection{Overall Search Method}
\begin{figure}[t]%
\centering%
\includegraphics[width=12cm]{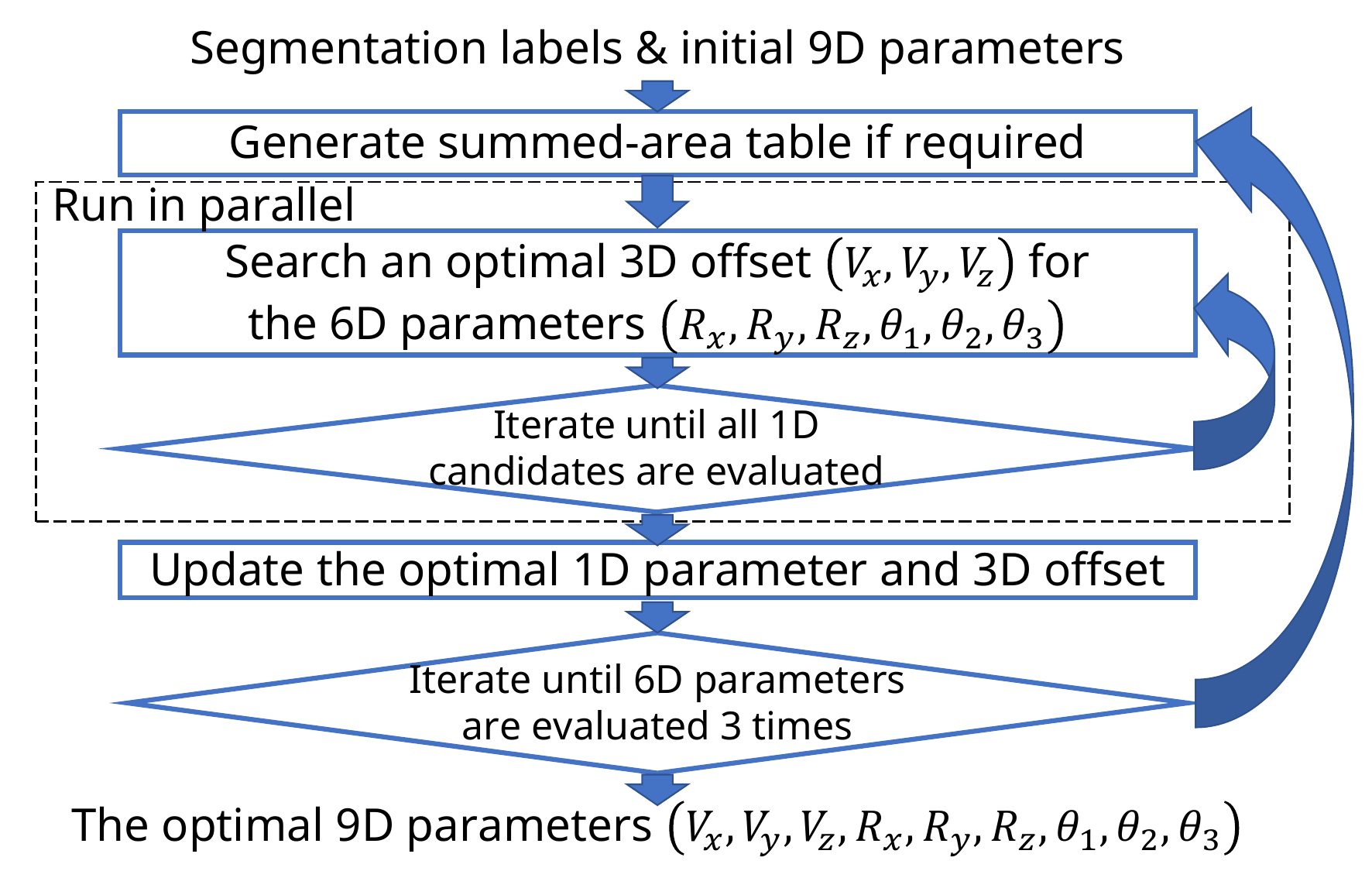}
\caption{A summary of the proposed search method.
The search of rectangular regions optimized 9D space using summed-area tables.
Summed-area tables were generated for individual 3D angles.
Whenever one of 6 parameters consisting of the 3D angle and 3D size was focused for searching,
remaining 5 parameters were not changed.
For each candidate of the focused parameter, the optimal 3D offset was also searched.
Each focused parameter was searched in parallel using multi-threading.
}\label{fig:9dsearch}
\end{figure}

As shown in Fig. \ref{fig:9dsearch},
the search of rectangular regions optimized 9-dimensional space using summed-area tables.
For individual 3D angles $\Theta$, summed-area tables were generated by
applying inverse matrices of the 3D rotation matrices to the tumor labels, and
generating tables using Eq. \ref{eq:summedareatable} for the rotated tumor labels.
The 3D size $R$ was searched without generating the tables again.

Whenever one of the 6 parameters $(R_x, R_y, R_z, \theta_1, \theta_2, \theta_3)$ was focused for searching,
the remaining 5 parameters were not changed.
For each candidate of the focused parameter, the 3D offset $V$ was also searched.
The focused parameter was searched using multi-threading
for utilizing multiple cores of a central processing unit (CPU).
In the search of the 3D offset $V$, the following two methods were implemented:
1D search which simulated the conventional method\cite{TumorVOIPlacement1}, and
3D full search which used an exhaustive search of the 3D offset $V$ for increasing preciseness.

Unless otherwise noted, the following parameters were used.
The minimum and maximum values of the elements of the 3D size $R$ were $5$ and $50$ mm, respectively.
In the cases of angle searches, the number of candidates were $9$.
In the first and remaining iterations, the angles were searched with the step size of $5$ and $5/9$ degrees, respectively.
The search centers were $0$ degrees for the first iteration and the results of the previous searches for remaining iterations.
The 3D offset $V$ was initialized to the centroid\cite{TumorVOIPlacement1} of the tumor labels.
The 3D size $R$ was initialized to the minimum values.
The 3D angle $\Theta$ was initialized to zeros.

\begin{table}[t]
\caption{Processing times, volume of rectangular regions and tumor fractions in the cases of target sizes of
(a) $15 \times 15 \times 15$ $\text{mm}^3$ and
(b) $30 \times 30 \times 30$ $\text{mm}^3$.
In the 1D cases, the proposed computation (p1D, 7 seconds) was slower than the conventional computation (c1D, 5-6 seconds).
In the 3D cases, the proposed computation (p3D, 8 seconds) was 100-500 times faster than the conventional computation (c3D, 11-40 minutes).
By comparing with the conventional metric (conv.m),
use of the proposed metric (prop.m) increased both volume of rectangular regions and tumor fractions without changing overall processing times.
Use of the segmentation network with 2 inputs (prop.m-2) slightly decreased tumor fractions.
}\label{table:eval1}%
\centering%
\small%
\begin{tabular}{|l|r|r|r|r|r|}
\multicolumn{6}{l}{(a) Evaluations with target size of $15 \times 15 \times 15$ $\text{mm}^3$ ($3375$ $\text{mm}^3$)}\\ \hline
& \multicolumn{3}{|c|}{Processing time (sec)} &  &  \\ \hline
& Segmentation & Search & Overall & Volume ($\text{mm}^3$) & Tumor (\%) \\ \hline
c1D+conv.m & $2.59 \pm 0.21$ & $2.66 \pm 0.09$ & $5.26 \pm 0.21$ & $3288.3 \pm 56.4$ & $81.66 \pm 13.98$ \\ \hline
p1D+conv.m & $2.53 \pm 0.18$ & $4.42 \pm 0.06$ & $6.95 \pm 0.20$ & $3286.7 \pm 52.2$ & $81.69 \pm 14.40$ \\ \hline
c3D+conv.m & $2.56 \pm 0.21$ & $672.60 \pm 6.07$ & $675.16 \pm 6.10$ & $3255.1 \pm 23.2$ & $84.05 \pm 11.12$ \\ \hline
p3D+conv.m & $2.60 \pm 0.09$ & $5.60 \pm 0.07$ & $8.19 \pm 0.12$ & $3255.1 \pm 23.2$ & $84.05 \pm 11.12$ \\ \hline
p3D+prop.m & $2.58 \pm 0.08$ & $5.73 \pm 0.08$ & $8.32 \pm 0.11$ & $3337.8 \pm 255.6$ & $96.88 \pm 13.37$ \\ \hline
p3D+prop.m-2 & $2.54 \pm 0.12$ & $5.78 \pm 0.07$ & $8.31 \pm 0.13$ & $3362.1 \pm 120.6$ & $96.33 \pm 14.30$ \\ \hline
\multicolumn{6}{l}{(b) Evaluations with target size of $30 \times 30 \times 30$ $\text{mm}^3$ ($27000$ $\text{mm}^3$)}\\ \hline
& \multicolumn{3}{|c|}{Processing time (sec)} &  &  \\ \hline
& Segmentation & Search & Overall & Volume ($\text{mm}^3$) & Tumor (\%) \\ \hline
c1D+conv.m & $2.63 \pm 0.19$ & $2.93 \pm 0.09$ & $5.56 \pm 0.21$ & $26950.0 \pm 0.0$ & $73.66 \pm 16.94$ \\ \hline
p1D+conv.m & $2.55 \pm 0.15$ & $4.45 \pm 0.07$ & $7.00 \pm 0.17$ & $26950.0 \pm 0.0$ & $73.69 \pm 16.86$ \\ \hline
c3D+conv.m & $2.58 \pm 0.21$ & $2408.03 \pm 13.90$ & $2410.61 \pm 13.91$ & $26950.0 \pm 0.0$ & $74.52 \pm 16.47$ \\ \hline
p3D+conv.m & $2.56 \pm 0.13$ & $5.74 \pm 0.09$ & $8.30 \pm 0.17$ & $26950.0 \pm 0.0$ & $74.52 \pm 16.47$ \\ \hline
p3D+prop.m & $2.56 \pm 0.11$ & $5.66 \pm 0.08$ & $8.23 \pm 0.14$ & $23231.0 \pm 7135.8$ & $92.38 \pm 11.73$ \\ \hline
p3D+prop.m-2 & $2.54 \pm 0.12$ & $5.70 \pm 0.08$ & $8.24 \pm 0.15$ & $23763.4 \pm 6525.8$ & $91.94 \pm 10.94$ \\ \hline
\end{tabular}
\end{table}
\begin{table}[tp]
\caption{Volumes of rectangular regions, tumor fractions, and size of rectangular region in all axes for various target conditions (*: $p<0.001$).
Results with various target sizes are shown in (a).
In the results with the conventional metric (conv.m),
$R_x$ were longer than those of $R_y$ and $R_z$.
In the results with the proposed metric (prop.m),
its sizes $R_x$, $R_y$ and $R_z$ were close to the target sizes $l_x$, $l_y$ and $l_z$.
Results with various target tumor fractions using target sizes of 20 and 40 mm are shown in (b) and (c), respectively.
The tumor fractions with the prop.m were higher than those with the conv.m in all evaluated cases.
In the cases of the conv.m, when target sizes were increased,
tumor fractions were increased without changing volumes of rectangular regions.
}\label{table:eval2}%
\centering%
\small%
\begin{tabular}{|l|r|r|r|r|r|} 
\multicolumn{6}{l}{(a) Evaluations with various target sizes (target tumor fraction: 90\%)}\\ \hline
& Volume ($\text{mm}^3$) & Tumor (\%) & $R_x$ (mm) & $R_y$ (mm) & $R_z$ (mm) \\ \hline
conv.m (10 mm) & $999.2 \pm 3.1$ & $85.16 \pm 11.47$ & $39.46 \pm 2.06$ & $5.08 \pm 0.31$ & $5.00 \pm 0.00$ \\ \hline
prop.m (10 mm) & $997.7 \pm 21.4$ & $97.67 \pm 10.95$* & $10.00 \pm 0.00$* & $10.00 \pm 0.00$* & $9.98 \pm 0.21$* \\ \hline
conv.m (15 mm) & $3255.1 \pm 23.2$ & $84.05 \pm 11.12$ & $49.91 \pm 0.42$ & $13.05 \pm 0.21$ & $5.00 \pm 0.00$ \\ \hline
prop.m (15 mm) & $3337.8 \pm 255.6$ & $96.88 \pm 13.37$* & $15.00 \pm 0.00$* & $14.99 \pm 0.11$* & $14.84 \pm 1.10$* \\ \hline
conv.m (20 mm) & $8000.0 \pm 0.0$ & $81.99 \pm 11.65$ & $50.00 \pm 0.00$ & $32.00 \pm 0.00$ & $5.00 \pm 0.00$ \\ \hline
prop.m (20 mm) & $7802.3 \pm 1017.4$ & $96.07 \pm 11.16$* & $19.97 \pm 0.24$* & $19.86 \pm 1.18$* & $19.57 \pm 2.31$* \\ \hline
conv.m (25 mm) & $15000.0 \pm 0.0$ & $74.74 \pm 15.95$ & $50.00 \pm 0.00$ & $50.00 \pm 0.00$ & $6.00 \pm 0.00$ \\ \hline
prop.m (25 mm) & $14464.1 \pm 2974.7$ & $94.35 \pm 11.90$* & $24.75 \pm 1.09$* & $24.38 \pm 2.74$* & $23.62 \pm 3.92$* \\ \hline
conv.m (30 mm) & $26950.0 \pm 0.0$ & $74.52 \pm 16.47$ & $49.00 \pm 0.00$ & $50.00 \pm 0.00$ & $11.00 \pm 0.00$ \\ \hline
prop.m (30 mm) & $23231.0 \pm 7135.8$* & $92.38 \pm 11.73$* & $29.20 \pm 2.81$* & $29.06 \pm 3.75$* & $26.67 \pm 6.74$* \\ \hline
\multicolumn{6}{l}{(b) Evaluations with various target tumor fractions (target size: $20 \times 20 \times 20$ $\text{mm}^3$)}\\ \hline
& Volume ($\text{mm}^3$) & Tumor (\%) & $R_x$ (mm) & $R_y$ (mm) & $R_z$ (mm) \\ \hline
conv.m (70\%) & $8000.0 \pm 0.0$ & $64.49 \pm 11.34$ & $50.00 \pm 0.00$ & $32.00 \pm 0.00$ & $5.00 \pm 0.00$ \\ \hline
prop.m (70\%) & $7848.6 \pm 871.8$ & $95.88 \pm 11.34$* & $19.94 \pm 0.38$* & $19.91 \pm 0.86$* & $19.68 \pm 1.92$* \\ \hline
conv.m (80\%) & $8000.0 \pm 0.0$ & $73.20 \pm 10.22$ & $50.00 \pm 0.00$ & $32.00 \pm 0.00$ & $5.00 \pm 0.00$ \\ \hline
prop.m (80\%) & $7848.6 \pm 871.8$ & $95.88 \pm 11.34$* & $19.94 \pm 0.38$* & $19.91 \pm 0.86$* & $19.68 \pm 1.92$* \\ \hline
conv.m (90\%) & $8000.0 \pm 0.0$ & $81.99 \pm 11.65$ & $50.00 \pm 0.00$ & $32.00 \pm 0.00$ & $5.00 \pm 0.00$ \\ \hline
prop.m (90\%) & $7778.4 \pm 1012.3$ & $96.11 \pm 11.15$* & $19.97 \pm 0.32$* & $19.90 \pm 0.86$* & $19.48 \pm 2.30$* \\ \hline
conv.m (100\%) & $8000.0 \pm 0.0$ & $88.17 \pm 14.82$ & $50.00 \pm 0.00$ & $32.00 \pm 0.00$ & $5.00 \pm 0.00$ \\ \hline
prop.m (100\%) & $7135.0 \pm 1937.9$* & $96.79 \pm 11.31$* & $19.85 \pm 0.99$* & $19.46 \pm 2.27$* & $18.02 \pm 4.37$* \\ \hline
\multicolumn{6}{l}{(c) Evaluations with various target tumor fractions (target size: $40 \times 40 \times 40$ $\text{mm}^3$)}\\ \hline
& Volume ($\text{mm}^3$) & Tumor (\%) & $R_x$ (mm) & $R_y$ (mm) & $R_z$ (mm) \\ \hline
conv.m (70\%) & $63700.0 \pm 0.0$ & $57.95 \pm 13.49$ & $49.00 \pm 0.00$ & $50.00 \pm 0.00$ & $26.00 \pm 0.00$ \\ \hline
prop.m (70\%) & $50572.6 \pm 20083.6$* & $81.06 \pm 12.58$* & $38.78 \pm 4.24$* & $39.15 \pm 4.03$* & $32.14 \pm 11.68$* \\ \hline
conv.m (80\%) & $63700.0 \pm 0.0$ & $63.18 \pm 16.90$ & $49.00 \pm 0.00$ & $50.00 \pm 0.00$ & $26.00 \pm 0.00$ \\ \hline
prop.m (80\%) & $45604.5 \pm 22343.8$* & $84.86 \pm 9.83$* & $38.83 \pm 4.19$* & $38.67 \pm 4.52$* & $28.94 \pm 13.25$ \\ \hline
conv.m (90\%) & $63700.0 \pm 0.0$ & $67.39 \pm 20.44$ & $49.00 \pm 0.00$ & $50.00 \pm 0.00$ & $26.00 \pm 0.00$ \\ \hline
prop.m (90\%) & $38734.5 \pm 24580.3$* & $90.29 \pm 7.18$* & $39.00 \pm 3.62$* & $35.82 \pm 7.80$* & $25.14 \pm 14.25$ \\ \hline
conv.m (100\%) & $63700.0 \pm 0.0$ & $70.07 \pm 23.03$ & $49.00 \pm 0.00$ & $50.00 \pm 0.00$ & $26.00 \pm 0.00$ \\ \hline
prop.m (100\%) & $26964.8 \pm 22412.0$* & $96.17 \pm 10.26$* & $37.20 \pm 5.00$* & $28.61 \pm 12.56$* & $20.66 \pm 12.43$* \\ \hline
\end{tabular}
\end{table}
\subsection{Evaluations}%


In the following evaluations, the validation dataset was used.
The processing times were measured on a CPU with 8 performance cores, 16 efficient cores and 32 processor threads.
The frequencies of the CPU were 3.2 GHz for the performance cores and 2.4 GHz for the efficient cores.
These cores were dynamically boosted up to 6.0 GHz.

As the first evaluations,
the improvement of the processing times using the summed-area tables were evaluated.
These evaluations measured processing times, volume of rectangular regions and tumor fractions.
The evaluated methods were
the conventional computation using Eq. \ref{eq:conventional} with the 1D search (c1D),
the conventional computation with the 3D full search (c3D),
the proposed computation using Eq. \ref{eq:summedarea} with the 1D search (p1D), and
the proposed computation with the 3D full search (p3D).
In the cases of prop.c3D, both the conventional metric Eq. \ref{eq:neglog} (conv.m) and 
proposed metric Eq. \ref{eq:proposed} (prop.m) were evaluated. The other cases were evaluated with conv.m only.
In these evaluations,
$l_x = l_y = l_z = 15$ and $30$ mm were evaluated.
In these evaluations, segmentation networks with both 4 and 2 inputs were evaluated for
simulating multi-contrast images after and before contrast enhancements, respectively.
The remaining parameters were
$f_{target} = 0.90$ and $\lambda_1 = 10^{-6}$.

As the second evaluations,
the differences of the search metrics were compared
by changing the target sizes, target tumor fractions, and $\lambda_2$.
The proposed method with 3-dimensional search was used for both
the conventional metric Eq. \ref{eq:neglog} (conv.m)
and proposed metric Eq. \ref{eq:proposed} (prop.m).
In these evaluations, the segmentation network with input channels of 4 was used.
When both conventional and proposed metrics were computed,
the Welch's t-test was used for computing $p$ values.

In the evaluations with various target sizes,
$l_x = l_y = l_z = 10, 15, 20, 25,$ and $30$ mm were evaluated.
The remaining parameters were
$f_{target} = 0.9$, $\lambda_1 = 10^{-6}$, $\lambda_2 = 0.01$ and $\beta = 0.1$.

In the evaluations with various target tumor fractions,
$f_{target} = 0.7, 0.8, 0.9,$ and $1.0$ were evaluated.
In these evaluations,
$l_x = l_y = l_z = 20$ and $40$ mm were evaluated.
The remaining parameters were
$\lambda_1 = 10^{-6}$, $\lambda_2 = 0.01$ and $\beta = 0.1$.

In the evaluations with various $\lambda_2$,
$\lambda_2 = 0.0001, 0.0002, 0.0005,
0.001, 0.002, 0.005, 0.01, 0.02, 0.05,$ and $0.1$ were evaluated.
The remaining parameters were
$f_{target} = 0.9$, $l_x = l_y = l_z = 40$ mm,
$\lambda_1 = 10^{-6}$, $\lambda_2 = 0.01$ and $\beta = 0.1$.

In addition, representative images were computed using both the 3-dimensional search and the proposed metric.
In this computation, the following conditions were used:
Target size of $160 \times 160 \times 15$ $\text{mm}^3$,
and segmentation network with 4 inputs.
This condition simulated a recommended condition of 2D MRSI\cite{MRSConsensus} after a contrast enhancement.

\section{Results}%
\begin{table}[t]
\caption{Results using the proposed metric with various $\lambda_2$.
As $\lambda_2$ increased, the volumes of rectangular regions got closer to the target volume ($64000$ $\text{mm}^3$).
As $\lambda_2$ decreased, the tumor fractions increased.
}\label{table:eval3}
\centering%
\begin{tabular}{|l|r|r|r|r|r|} \hline
$\lambda_2$& Volume ($\text{mm}^3$) & Tumor (\%) & $R_x$ (mm) & $R_y$ (mm) & $R_z$ (mm) \\ \hline
0.0001 & $22703.6 \pm 21098.2$ & $96.52 \pm 10.58$ & $37.21 \pm 5.37$ & $28.29 \pm 12.88$ & $17.00 \pm 11.93$ \\ \hline
0.0002 & $26964.8 \pm 22412.0$ & $96.17 \pm 10.26$ & $37.20 \pm 5.00$ & $28.61 \pm 12.56$ & $20.66 \pm 12.43$ \\ \hline
0.0005 & $35287.3 \pm 23480.0$ & $93.53 \pm 10.56$ & $36.59 \pm 5.33$ & $31.90 \pm 10.99$ & $25.97 \pm 12.41$ \\ \hline
0.001 & $39618.6 \pm 23567.7$ & $90.00 \pm 10.81$ & $37.99 \pm 5.01$ & $35.23 \pm 8.73$ & $26.60 \pm 12.85$ \\ \hline
0.002 & $38734.5 \pm 24580.3$ & $90.29 \pm 7.18$ & $39.00 \pm 3.62$ & $35.82 \pm 7.80$ & $25.14 \pm 14.25$ \\ \hline
0.005 & $40210.6 \pm 23658.1$ & $89.99 \pm 7.43$ & $38.89 \pm 3.36$ & $34.39 \pm 9.47$ & $27.30 \pm 12.99$ \\ \hline
0.01 & $49513.2 \pm 19570.8$ & $83.44 \pm 15.74$ & $36.71 \pm 6.40$ & $37.63 \pm 6.15$ & $33.57 \pm 9.73$ \\ \hline
0.02 & $64000.0 \pm 0.0$ & $70.09 \pm 23.42$ & $40.00 \pm 0.00$ & $40.00 \pm 0.00$ & $40.00 \pm 0.00$ \\ \hline
0.05 & $64000.0 \pm 0.0$ & $70.21 \pm 23.18$ & $40.00 \pm 0.00$ & $40.00 \pm 0.00$ & $40.00 \pm 0.00$ \\ \hline
0.1 & $64000.0 \pm 0.0$ & $70.21 \pm 23.18$ & $40.00 \pm 0.00$ & $40.00 \pm 0.00$ & $40.00 \pm 0.00$ \\ \hline
\end{tabular}
\end{table}
\begin{figure}[t]%
\centering%
\includegraphics[width=15cm]{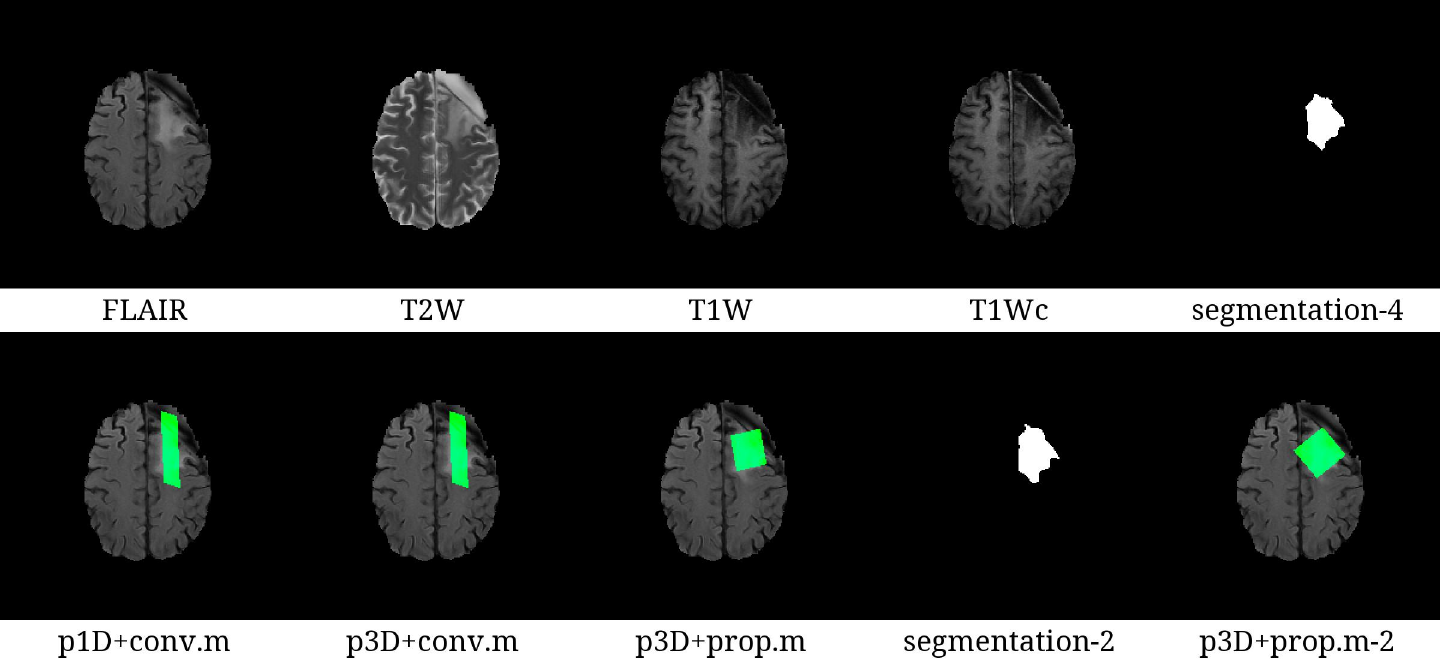}
\caption{%
Representative images, tumor labels, and rectangular regions in the cases of searching $30 \times 30 \times 30$ $\text{mm}^3$ regions.
The segmentation images represent estimated tumor regions.
The rectangular regions were filled with green.
As shown in the cases of the rectangular regions using the conventional metric (conv.m),
the conventional metric preferred oblongs.
The extracted shapes of the rectangular regions were similar in the cases of
both 1-dimensional search (p1D) and 3-dimensional search (p3D).
The shapes of the rectangular regions using the proposed metric (prop.m) were close to cubes
since the proposed metric gave priority to cubes over oblongs.
When the segmentation network with 2 inputs was used instead of that with 4 inputs,
the results were slightly changed as shown in segmentation-2 and prop.m-2.
}\label{fig:examplesinglevoxel}
\end{figure}
\begin{figure}[t]%
\centering%
\includegraphics[width=12cm]{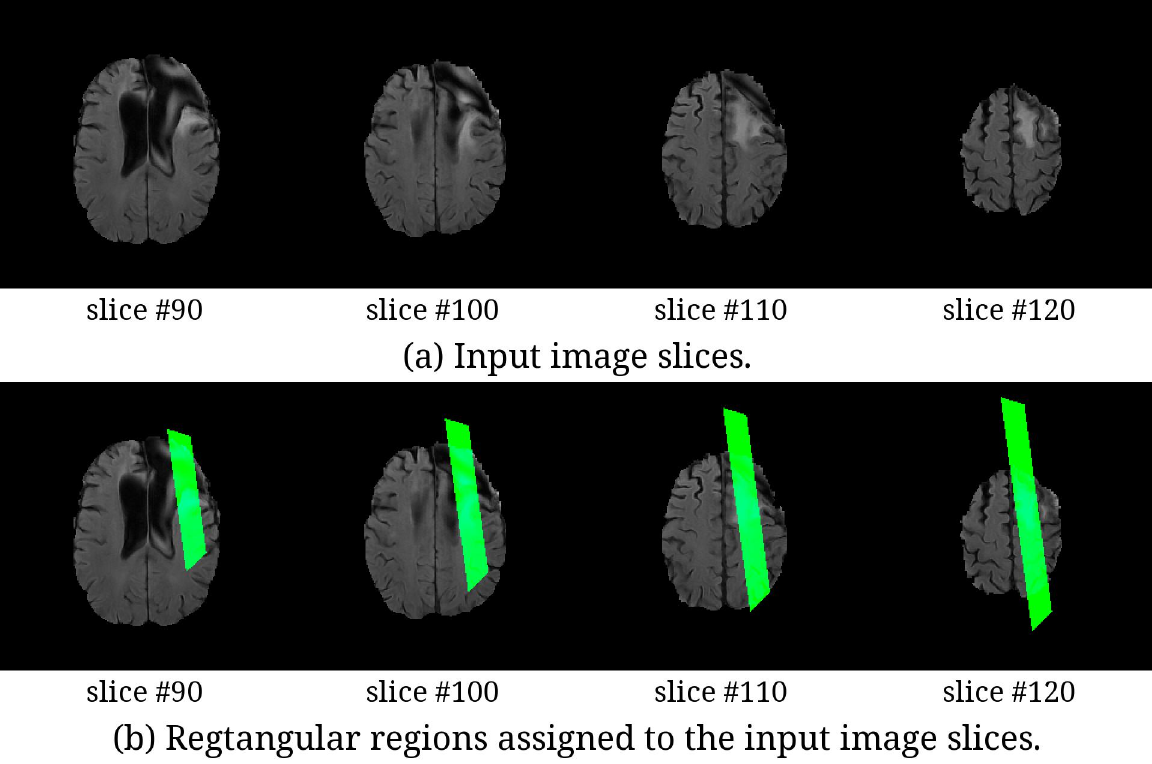}
\caption{%
Representative images and rectangular regions in the cases of searching $120 \times 120 \times 15$ $\text{mm}^3$ regions.
The rectangular regions were filled with green.
Both the 3-dimensional search and the proposed metric were used for finding the rectangular regions.
The rectangular region was rotated 3-dimensionally for putting the region on tumors from 90th to 120th slices.
The region size was exactly same as the target size.
}\label{fig:examplemrsi}
\end{figure}

The processing times, volume of rectangular regions and tumor fractions are shown in Table \ref{table:eval1}.
In the 1D cases, the proposed computation (7 seconds) was slower than the conventional computation (5-6 seconds).
In the 3D cases, the proposed computation (8 seconds) was 100-500 times faster than the conventional computation (11-40 minutes).
By comparing with the conv.m,
use of the prop.m increased both volume of rectangular regions and tumor fractions without changing overall processing times.
In the cases of searching $3375$ $\text{mm}^3$ rectangular regions, the processing times of the conventional computation were
$5.26 \pm 0.21$ and $675.16 \pm 6.10$ seconds for the 1D and 3D full search, respectively.
Those of the proposed computation were
$6.95 \pm 0.20$ and $8.19 \pm 0.12$ seconds for the 1D and 3D full search, respectively.
In the cases of searching $27000$ $\text{mm}^3$ rectangular regions, the processing times of the conventional computation were
$5.56 \pm 0.21$ and $2410.6 \pm 13.9$ seconds for the 1D and 3D full search, respectively.
Those of the proposed computation were
$7.00 \pm 0.17$ and $8.30 \pm 0.17$ seconds for the 1D and 3D full search, respectively.
Representative images and their rectangular regions are shown
in Fig. \ref{fig:examplesinglevoxel}.
As shown in the cases of the rectangular regions using the conventional metric (conv.m),
the conventional metric preferred oblongs.
In contrast, the shapes of the rectangular regions using the proposed metric were close to cubes
since the proposed metric gave priority to cubes over oblongs.

The volumes of rectangular regions, tumor fractions, and size of rectangular region for various target conditions
are shown in Table \ref{table:eval2}.
In Table \ref{table:eval2},
items with significant differences between the conv.m and prop.m using the Welch's t-test are marked with asterisks.

The results with various target sizes are shown in Table \ref{table:eval2} (a).
In the results with the conv.m, $R_x$ were longer than those of $R_y$ and $R_z$.
In the results with the prop.m,
its sizes $R_x$, $R_y$ and $R_z$ were close to the target sizes $l_x$, $l_y$ and $l_z$.
The results with various target tumor fractions using two target sizes are shown in Table \ref{table:eval2} (b) and (c).
The tumor fractions with the prop.m were higher than those with the conv.m in all evaluated cases.
In the cases of the conv.m, as the target tumor fraction increased,
the tumor fraction increased without changing the volume of the rectangular region.

The results using the proposed metric with various $\lambda_2$ are shown in Table \ref{table:eval3}.
As $\lambda_2$ increased, the volume of the rectangular region got closer to the target volume.
As $\lambda_2$ decreased, the tumor fraction increased.
The balance between the volumes and tumor fractions was changed in the cases of $\lambda_2 \le 0.02$.

Representative images which simulated a recommended condition of 2D MRSI are shown
in Fig. \ref{fig:examplemrsi}.
The rectangular region was rotated 3-dimensionally for putting the region on tumors from 90th to 120th slices.
The region size was exactly same as the target size.

\section{Discussion}

The results demonstrated the effectiveness of the proposed which searched rectangular regions precisely in 8 seconds.
As shown in Table \ref{table:eval1}, the proposed computation reduced processing times of the 3D full search.
In addition, the proposed metric improved the quality of the rectangular regions as shown in Table \ref{table:eval2} and Fig. \ref{fig:examplesinglevoxel}.
As shown in Table \ref{table:eval3},
the balance between the volumes and tumor fractions could be changed by controlling $\lambda_2$.
While evaluations were limited to the BraTS dataset, the proposed method is promising for implementing fast and precise search of rectangular regions.

Use of the summed-area tables was efficient for the 3D full search as shown in Table \ref{table:eval1}.
In most cases, the rectangular regions with the 1D search were similar to the rectangular regions with the 3D full search
since the 3D offset was initialized to its centroid.
However, this initialization relied on the assumption that there was only one tumor region.
As a potential problem, in the cases of two or more isolated tumor regions,
the 3D offset can be initialized at the location far from large tumor regions.
In such cases, the 1D search cannot find optimal rectangular regions
since the 1D search cannot put candidates at large tumor regions.
The 3D full search can avoid this potential problem since it does not need initialization.

The proposed metric was effective for improving shapes of the rectangular regions and increasing tumor fractions.
In Table \ref{table:eval2} (b-c),
it was shown that the conv.m could increase the tumor fraction without changing the volume of the rectangular region.
This behavior means that the conv.m reduced tumor fractions by moving rectangular regions to outsides
if the tumor fractions were greater than the target tumor fraction.
In contrast, in the cases of the proposed method,
the tumor fractions were higher than the target tumor fractions
except but the cases of 100\%.
Since there were no such unexpected behaviors in the cases of the proposed metric,
the first term of Eq. \ref{eq:proposed} was effective for assigning better rectangular regions whenever possible.
The shapes of the rectangular regions assigned by the conventional metric were oblongs.
Since the rectangular regions assigned by the proposed metric were close to cubes,
the second term of Eq. \ref{eq:proposed} was effective for avoiding oblong rectangular regions.
%

The remained work is to evaluate the proposed method with SV-MRS and MRSI scans of brain tumors.
The proposed method was evaluated with a publicly available dataset only
since there were no environments for evaluating the proposed method.

Applying the proposed method with other applications to non-tumor SV-MRS and MRSI scans are also remained as future work.
The proposed method could be used for other applications by changing segmentation networks
since the proposed method did not use ground-truth rectangular regions in learning neural networks\cite{TumorVOIPlacement2}.

\section{Conclusion}

The proposed method is promising for implementing fast and precise search of rectangular regions.
In the proposed method,
the proposed computation reduced processing times of the 3D full search, and
the proposed metric improved the quality of the assigned rectangular regions.

\section{Conflicts of Interest}

Hidenori Takeshima and Shuki Maruyama are employees of Canon Medical Systems Corporation.


\bibliographystyle{elsarticle-num}
\bibliography{paperref}

\begin{thebibliography}{10}
\expandafter\ifx\csname url\endcsname\relax
  \def\url#1{\texttt{#1}}\fi
\expandafter\ifx\csname urlprefix\endcsname\relax\def\urlprefix{URL }\fi
\expandafter\ifx\csname href\endcsname\relax
  \def\href#1#2{#2} \def\path#1{#1}\fi

\bibitem{MRITumorAnalysis}
S.~Bauer, R.~Wiest, L.-P. Nolte, M.~Reyes, A survey of mri-based medical image
  analysis for brain tumor studies, Physics in Medicine \& Biology 58~(13)
  (2013) R97--R129.
\newblock \href {https://doi.org/10.1088/0031-9155/58/13/R97}
  {\path{doi:10.1088/0031-9155/58/13/R97}}.

\bibitem{MRSTumorAnalysis}
B.~D. Weinberg, M.~Kuruva, H.~Shim, M.~E. Mullins, Clinical applications of
  magnetic resonance spectroscopy in brain tumors: From diagnosis to treatment,
  Radiologic Clinics of North America 59~(3) (2021) 349--362.
\newblock \href {https://doi.org/10.1016/j.rcl.2021.01.004}
  {\path{doi:10.1016/j.rcl.2021.01.004}}.

\bibitem{MRSConsensus}
M.~Wilson, O.~Andronesi, P.~B. Barker, R.~Bartha, A.~Bizzi, P.~J. Bolan,
  et~al., Methodological consensus on clinical proton {MRS} of the brain:
  Review and recommendations, Magnetic Resonance in Medicine 82~(2) (2019)
  527--550.
\newblock \href {https://doi.org/10.1002/mrm.27742}
  {\path{doi:10.1002/mrm.27742}}.

\bibitem{MRSGABA}
N.~A. Puts, R.~A. Edden, In vivo magnetic resonance spectroscopy of {GABA}: A
  methodological review, Progress in Nuclear Magnetic Resonance Spectroscopy 60
  (2012) 29--41.
\newblock \href {https://doi.org/10.1016/j.pnmrs.2011.06.001}
  {\path{doi:10.1016/j.pnmrs.2011.06.001}}.

\bibitem{MRS2HG}
C.~H. Suh, H.~S. Kim, S.~C. Jung, C.~G. Choi, S.~J. Kim, 2-hydroxyglutarate
  {MR} spectroscopy for prediction of isocitrate dehydrogenase mutant glioma: a
  systemic review and meta-analysis using individual patient data,
  Neuro-Oncology 20~(12) (2018) 1573--1583.
\newblock \href {https://doi.org/10.1093/neuonc/noy113}
  {\path{doi:10.1093/neuonc/noy113}}.

\bibitem{AnatomicalVOIPlacement1}
W.~Dou, O.~Speck, T.~Benner, J.~Kaufmann, M.~Li, K.~Zhong, M.~Walter, Automatic
  voxel positioning for {MRS} at 7 {T}, Magnetic Resonance Materials in
  Physics, Biology and Medicine 28 (2015) 259--270.
\newblock \href {https://doi.org/10.1007/s10334-014-0469-9}
  {\path{doi:10.1007/s10334-014-0469-9}}.

\bibitem{AnatomicalVOIPlacement2}
Y.~W. Park, D.~K. Deelchand, J.~M. Joers, B.~Hanna, A.~Berrington, J.~S.
  Gillen, et~al., {AutoVOI}: real-time automatic prescription of
  volume-of-interest for single voxel spectroscopy, Magnetic Resonance in
  Medicine 80~(5) (2018) 1787--1798.
\newblock \href {https://doi.org/10.1002/mrm.27203}
  {\path{doi:10.1002/mrm.27203}}.

\bibitem{AnatomicalVOIPlacement3}
J.~H. Bishop, A.~Geoly, N.~Khan, C.~Tischler, R.~Krueger, P.~Keshava, H.~Amin,
  L.~Baltusis, H.~Wu, D.~Spiegel, N.~Williams, M.~D. Sacchet, Real-time
  semi-automated and automated voxel placement using {fMRI} targets for
  repeated acquisition magnetic resonance spectroscopy, Journal of Neuroscience
  Methods 392 (2023) 109853.
\newblock \href {https://doi.org/10.1016/j.jneumeth.2023.109853}
  {\path{doi:10.1016/j.jneumeth.2023.109853}}.

\bibitem{TumorVOIPlacement1}
P.~J. Bolan, F.~Branzoli, A.~L. Di~Stefano, L.~Nichelli, R.~Valabregue, S.~L.
  Saunders, et~al., Automated acquisition planning for magnetic resonance
  spectroscopy in brain cancer, Medical Image Computing and Computer Assisted
  Intervention (MICCAI), Springer International Publishing, Cham, 2020, pp.
  730--739.
\newblock \href {https://doi.org/10.1007/978-3-030-59728-3_71}
  {\path{doi:10.1007/978-3-030-59728-3_71}}.

\bibitem{TumorVOIPlacement2}
S.~Lee, F.~Branzoli, T.~Nguyen, O.~Andronesi, A.~Lin, R.~Liserre, et~al., A
  deep learning approach for placing magnetic resonance spectroscopy voxels in
  brain tumors, Medical Image Computing and Computer Assisted Intervention
  (MICCAI), Springer Nature Switzerland, Cham, 2024, pp. 543--552.
\newblock \href {https://doi.org/10.1007/978-3-031-72384-1_51}
  {\path{doi:10.1007/978-3-031-72384-1_51}}.

\bibitem{blockmatchingsurvey}
Y.-W. Huang, C.-Y. Chen, C.-H. Tsai, C.-F. Shen, L.-G. Chen, Survey on block
  matching motion estimation algorithms and architectures with new results,
  Journal of VLSI signal processing systems for signal, image and video
  technology 42~(3) (2006) 297--320.
\newblock \href {https://doi.org/10.1007/s11265-006-4190-4}
  {\path{doi:10.1007/s11265-006-4190-4}}.

\bibitem{BlochMatchingFFT}
S.~Kilthau, M.~Drew, T.~Moller, Full search content independent block matching
  based on the fast {Fourier} transform, Proceedings of International
  Conference on Image Processing, 2002, pp. I--669--672.
\newblock \href {https://doi.org/10.1109/ICIP.2002.1038113}
  {\path{doi:10.1109/ICIP.2002.1038113}}.

\bibitem{ActiveSearch}
V.~Vinod, H.~Murase, Focused color intersection with efficient searching for
  object extraction, Pattern Recognition 30~(10) (1997) 1787--1797.
\newblock \href {https://doi.org/10.1016/S0031-3203(96)00192-6}
  {\path{doi:10.1016/S0031-3203(96)00192-6}}.

\bibitem{SummedAreaTables}
F.~C. Crow, Summed-area tables for texture mapping, Proceedings of the 11th
  Annual Conference on Computer Graphics and Interactive Techniques,
  Association for Computing Machinery, New York, NY, USA, 1984, pp. 207--212.
\newblock \href {https://doi.org/10.1145/800031.808600}
  {\path{doi:10.1145/800031.808600}}.

\bibitem{IntegralImage}
P.~Viola, M.~J. Jones, Robust real-time face detection, International Journal
  of Computer Vision 57 (2004) 137--154.
\newblock \href {https://doi.org/10.1023/B:VISI.0000013087.49260.fb}
  {\path{doi:10.1023/B:VISI.0000013087.49260.fb}}.

\bibitem{IntegralImageHighDim}
E.~Tapia, A note on the computation of high-dimensional integral images,
  Pattern Recognition Letters 32~(2) (2011) 197--201.
\newblock \href {https://doi.org/10.1016/j.patrec.2010.10.007}
  {\path{doi:10.1016/j.patrec.2010.10.007}}.

\bibitem{Takeshima1}
H.~Takeshima, S.~Maruyama, A fast 3-dimensional full search algorithm for
  setting volume of interests of {MR} spectroscopy in brain tumor images,
  Proceedings of the 2025 ISMRM \& ISMRT Annual Meeting, 2025, p. 4830.

\bibitem{Takeshima2}
H.~Takeshima, S.~Maruyama, Automatic placement of volume-of-interest for
  magnetic resonance spectroscopy with flexible search criteria in brain tumor
  images (in {Japanese}), Proceedings of the Japanese Society for Magnetic
  Resonance in Medicine (JSMRM), 2025, pp. PS9--1.

\bibitem{BRATS1}
B.~H. Menze, A.~Jakab, S.~Bauer, J.~Kalpathy-Cramer, K.~Farahani, J.~Kirby,
  et~al., The multimodal brain tumor image segmentation benchmark ({BRATS}),
  IEEE Transactions on Medical Imaging 34~(10) (2015) 1993--2024.
\newblock \href {https://doi.org/10.1109/TMI.2014.2377694}
  {\path{doi:10.1109/TMI.2014.2377694}}.

\bibitem{BRATS2}
S.~Bakas, H.~Akbari, A.~Sotiras, M.~Bilello, M.~Rozycki, J.~S. Kirby, J.~B.
  Freymann, K.~Farahani, C.~Davatzikos, Advancing the cancer genome atlas
  glioma {MRI} collections with expert segmentation labels and radiomic
  features, Scientific Data 4 (2017) 170117.
\newblock \href {https://doi.org/10.1038/sdata.2017.117}
  {\path{doi:10.1038/sdata.2017.117}}.

\bibitem{BRATS3}
S.~Bakas, M.~Reyes, A.~Jakab, S.~Bauer, M.~Rempfler, A.~Crimi, et~al.,
  Identifying the best machine learning algorithms for brain tumor
  segmentation, progression assessment, and overall survival prediction in the
  {BRATS} challenge, arXiv preprint 1811.02629 (2019).
\newblock \href {https://doi.org/10.48550/arXiv.1811.02629}
  {\path{doi:10.48550/arXiv.1811.02629}}.

\bibitem{BRATS4}
S.~Bakas, H.~Akbari, A.~Sotiras, M.~Bilello, M.~Rozycki, J.~Kirby, J.~Freymann,
  K.~Farahani, C.~Davatzikos, Segmentation labels for the pre-operative scans
  of the {TCGA}-{GBM} collection [data set], The Cancer Imaging Archive (2017).
\newblock \href {https://doi.org/10.7937/K9/TCIA.2017.KLXWJJ1Q}
  {\path{doi:10.7937/K9/TCIA.2017.KLXWJJ1Q}}.

\bibitem{BRATS5}
S.~Bakas, H.~Akbari, A.~Sotiras, M.~Bilello, M.~Rozycki, J.~Kirby, J.~Freymann,
  K.~Farahani, C.~Davatzikos, Segmentation labels and radiomic features for the
  pre-operative scans of the {TCGA}-{LGG} collection [data set], The Cancer
  Imaging Archive (2017).
\newblock \href {https://doi.org/10.7937/K9/TCIA.2017.GJQ7R0EF}
  {\path{doi:10.7937/K9/TCIA.2017.GJQ7R0EF}}.

\bibitem{UNet}
O.~Ronneberger, P.~Fischer, T.~Brox, U-{N}et: Convolutional networks for
  biomedical image segmentation, Medical Image Computing and Computer-Assisted
  Intervention (MICCAI), Springer International Publishing, Cham, 2015, pp.
  234--241.
\newblock \href {https://doi.org/10.1007/978-3-319-24574-4_28}
  {\path{doi:10.1007/978-3-319-24574-4_28}}.

\bibitem{EfficientNet}
M.~Tan, Q.~V. Le, {EfficientNet}: Rethinking model scaling for convolutional
  neural networks, arXiv preprint 1905.11946 (2019).
\newblock \href {https://doi.org/10.48550/arXiv.1905.11946}
  {\path{doi:10.48550/arXiv.1905.11946}}.

\bibitem{SegmentationModelsPytorch}
P.~Iakubovskii, Segmentation models pytorch, accessed September 29, 2025.
  \url{https://github.com/qubvel/segmentation_models.pytorch}, GitHub
  repository. (2019).

\bibitem{MnasNet}
M.~Tan, B.~Chen, R.~Pang, V.~Vasudevan, M.~Sandler, A.~Howard, Q.~V. Le,
  {MnasNet}: Platform-aware neural architecture search for mobile, arXiv
  preprint 1807.11626 (2018).
\newblock \href {https://doi.org/10.48550/arXiv.1807.11626}
  {\path{doi:10.48550/arXiv.1807.11626}}.

\bibitem{SiLU}
S.~Elfwing, E.~Uchibe, K.~Doya, Sigmoid-weighted linear units for neural
  network function approximation in reinforcement learning, arXiv preprint
  1702.03118 (2017).
\newblock \href {https://doi.org/10.48550/arXiv.1702.03118}
  {\path{doi:10.48550/arXiv.1702.03118}}.

\bibitem{Adam}
D.~P. Kingma, J.~Ba, Adam: A method for stochastic optimization, arXiv preprint
  1412.6980 (2017).
\newblock \href {https://doi.org/10.48550/arXiv.1412.6980}
  {\path{doi:10.48550/arXiv.1412.6980}}.

\end{thebibliography}

\end{document}